\documentstyle[11pt]{article}
\catcode`\@=11
\begin{document} \openup6pt

\title{INFLATION IN BIANCHI MODELS AND THE COSMIC NO
HAIR THEOREM IN  BRANE WORLD}

\author{Bikash Chandra Paul\thanks{Electronic mail : bcpaul@iucaa.ernet.in}\\
	Physics Department, North Bengal University, \\
Siliguri, Dist. : Darjeeling, Pin : 734 430, West Bengal, India }

\date{}
\maketitle
\vspace{0.5in}

\begin{abstract}
In this paper, the cosmic no hair theorem for anisotropic Bianchi models which 
admit inflation with a scalar field is studied in the framework of Brane
 world. It is found that  all Bianchi models 
except Bianchi type IX, transit to an inflationary regime with 
vanishing anisotropy. In the Brane world, anisotropic universe approaches the
inflationary era much faster than  that in the general theory of relativity. 
The form of the potential does not affect the evolution in the inflationary epoch.
However, the late time behaviour is controlled by a constant additive factor in the  potential for the inflaton field.
\end{abstract}
 
\vspace{0.2cm}

PACS number(s) : 04.50.+h, 98.80.Cq

\pagebreak

Over the past couple of years there is a growing interest to study 
cosmological models in the framework of higher dimensional space-time
motivated by  the developments in 
superstring and M-theory  [1,2].
In these theories, gravity is a higher dimensional theory which reduces 
effectively 4-dimensional at lower energy scale. Such higher dimensional
 theories open up the possibility 
of solving the hierarchy problem in particle physics by considering  large 
compactified extra dimensions and make the string
 scale accessible to the future accelerators [3].  In such scenario our 
observed 
universe is described by a brane embedded in higher dimensional spacetime 
[4] and usual matter field and force except for gravity that are confined on 
the brane. 
The gravitational field may
 propagate through the bulk dimensions perpendicular to the brane. 
Randall and Sundrum [2] shown that even if the extra dimensions are not 
compact, four dimensional Newtonian gravity is recovered in five dimensional 
anti-de Sitter spacetime ($AD$$S_{5}$) in low energy limit.

Recently in the brane world scenario, homogeneous and isotropic cosmological
 models are studied [5] which describe the early universe satisfactorily. 
  Maartens { \it et al.} [6]  shown that  chaotic 
inflation can be accommodated on the brane and found that the
modified braneworld  Friedmann equation leads to a stronger condition for 
inflation. The brane effects ease the 
condition for slow-roll inflation 
for  a given potential. 
 Maartens, Sahni
and Saini [7] also explored the  behavior of an anisotropic Bianchi type-I 
brane world in the
 presence of a scalar field. They found that a large anisotropy on the 
 brane does not prevent 
inflation, moreover, a large anisotropy
 enhances more damping into the scalar field equation of motion, 
resulting greater inflation. In brane world, cosmological solutions  with
 a singularity and
without singularity in an anisotropic Bianchi type-I universe 
are obtained by the author [8]  which accommodate inflationary regime.
Toporensky [9] explored the shear dynamics
in Bianchi type I cosmological 
model on a brane with perfect fluid obeying the equation of state $p = (\gamma - 1) \rho$ where $\rho$ and $p$ are energy density and pressure respectively and $\gamma $ is a constant.  He found that for $1 <  \gamma  < 2$, the shear attains a maximum value during 
its transition from non-standard to standard cosmology i.e., when the 
matter energy density is comparable to the brane tension.
 Campos and Sopuerta [10] studied qualitatively   two of the  Bianchi universes, Bianchi-I and Bianchi
type V 
   in the Randall-Sundrum brane world scenario with 
matter on the brane obeying the barotropic  equation of state. It is found that anisotropic Bianchi I and V braneworlds 
always isotropize,  although there could be intermediate stages in which
the anisotropy grows. The brane world scenario near the bigbang is found to differ
from the general theory of  relativity ( henceforth, GTR ). The 
anisotropy dominates for  $ \gamma  \leq 1 $ in the braneworld whereas in GTR
it happens for 
$ \gamma  \leq 2 $.  Frolov [11] carried out geometrical construction of
 the Randall-Sundrum braneworld, 
without the assumption of spatial 
isotropy but by considering a homogeneous and 
anisotropic Kasner type solution 
of the Einstein-Ads equation in the bulk. Recently, Santos {\it et al.} [12] 
studied  no hair theorem for global anistropy in the brane world 
scenario following Wald [13]. It is found that the brane matter and bulk metric in the anisotropic brane
under certain condition evolves  asymptotically to a
 de Sitter space-time in the presence of a positive four dimensional cosmological constant. 
In this brief report
cosmic no hair
 theorem in anistropic Bianchi brane in the presence of matter described by a scalar field is studied. 
In the brane the effective four dimensional cosmological constant may be assumed to be
 zero by the choice of the brane tension which is also considered here.

Consider a five dimensional (bulk) space-time in which the Einstein's field equation is
given by 
\begin{equation}
G_{AB}^{(5)} = \tilde{\kappa}^{2} \left[ - g_{AB}^{(5)} \Lambda_{(5)} + 
 T_{AB}^{(5)} \right] 
\end{equation}
with $T_{AB}^{(5)} = \delta (y) [ - \lambda g_{AB} + T_{AB} ]$. 
Here $\tilde{\kappa}$ represents the  five dimensional gravitational 
coupling constant, $g_{AB}^{(5)}$, 
$G_{AB}^{(5)} $ and  $\Lambda_{(5)} $ are the metric, Einstein tensor and 
the cosmological constant of the bulk space-time respectively, $T_{AB}$ is 
the matter energy momentum tensor. We have  $\tilde{\kappa} = 
\frac{8 \pi}{M_{P}^{3}}$, where $M_{P} = 1.2 
\times 10^{19} $ GeV.  A natural  choice of coordinates is $x^{A} = ( x^{\mu},
 y ) $ where $x^{\mu} = (t, x^{i})$ are space-time coordinates on the brane. 
The the upper case Latin letters
($A, B, ... = 0, ..., 4$)  represents  coordinate indices in 
the bulk spacetime, the Greek letters  ($\mu, \nu, ... = 0, ..., 3$)
for the coordinate indices in the four dimensional
spacetime and the
small case latin letters  ($i, j = 1, 2, 3 $) for three space. 
The space-like hypersurface $x^{4} = y = 0$ gives the brane world and
 $g_{AB}$ is its induced metric, $\lambda $ is the tension of the brane 
which is assumed to be positive in order to recover conventional general
theory of gravity (GTR) on the brane. The bulk cosmological constant 
$\Lambda_{(5)} $ is negative and represents the five dimensional cosmological
constant.

The field equations induced on the brane are derived using geometric 
approach [14] leading to new terms which carry bulk effects on the
 brane. The modified dynamical equations on the brane is
\begin{equation}
G_{\mu \nu}^{(5)} = - \Lambda g_{\mu \nu} +   \kappa^{2} T_{\mu \nu} 
 + \tilde{\kappa}^{4} S_{\mu \nu} - E_{\mu \nu}^{(5)} .
\end{equation}
The effective cosmological 
constant $\Lambda$ and the four dimensional constant $\kappa$ 
 on the brane are given by
\[
\Lambda = \frac{|\Lambda_{5}|}{2}  \left[ 
\left( \frac{\lambda}{\lambda_{c}} \right)^{2} - 1 \right] 
\]
\begin{equation}
\kappa^{2}  = \frac{1}{6} \lambda \; \tilde{\kappa}^{4},
\end{equation}
respectively, where $\lambda_{c}$ is the critical brane tension which is 
given by 
\begin{equation}
\lambda_{c} = 6 \frac{|\Lambda_{5}|}{\kappa_{5}^{2}} .
\end{equation}
However one can make the effective four dimensional cosmological 
constant zero by a choice of the brane tension. The 
extra dimensional corrections to the 
Einstein equations on the brane are of two types and are given by : 

$\bullet $
$S_{\mu \nu} $ :  quadratic in the matter variables which is
\begin{equation}
S_{\mu \nu} =  \frac{1}{12} T T_{\mu \nu} - \frac{1}{4}  T_{\mu \alpha} 
T^{\alpha}_{\nu} + \frac{1}{24} g_{\mu \nu} \left[ 3 T_{\alpha \beta} 
T_{\alpha \beta} - (T^{\alpha}_{\alpha})^{2} \right] .
\end{equation}
where $T = T^{\alpha}_{\alpha}$, $S_{\mu \nu}$ is significant at high 
energies i.e., $\rho > \lambda$, 

$\bullet$
$E_{\mu \nu}^{(5)} $ :  occurs  due to the non-local effects from the 
free gravitational field in the bulk, which  enters  in the equation via the
 projection 
${\bf E}^{(5)}_{AB} =  C^{(5)}_{ACBD} n^{C} n^{D}$ where $n^{A}$ is 
normal to the surface ($n^{A} n_{A} = 1$). The term is symmetric and 
traceless 
and without components orthogonal to the brane, so ${\bf E}_{AB} n^{B} = 0$
and ${\bf E}_{AB} \rightarrow  {\it E}_{\mu \nu} g^{\mu}_{A} g^{\nu}_{B} $
as $y \rightarrow 0$.

To anlyse the cosmological evolution, we consider two components of the
 dynamical
 equation (2). First we consider the "initial-value" constraint equation
\begin{equation}
G_{\mu \nu} n^{\mu}  n^{\nu} =  \kappa^{2} T_{\mu \nu}  n^{\mu} n^{\nu}
+ \tilde{\kappa}^{4}  S_{\mu \nu} n^{\mu} n^{\nu} - {\it E}_{\mu \nu} n^{\mu} 
n^{\nu} 
\end{equation}
and the Raychaudhuri equation
\begin{equation}
R_{\mu \nu} n^{\mu}  n^{\nu} =  \kappa^{2} \left( T_{\mu \nu}  - \frac{1}{2} 
g_{\mu \nu} T \right) n^{\mu} n^{\nu}
+ \tilde{\kappa}^{4} \left( S_{\mu \nu} - \frac{1}{2} g_{\mu \nu} S \right)
 n^{\mu} n^{\nu} - {\it E}_{\mu \nu} n^{\mu} 
n^{\nu} 
\end{equation}
where $ n^{\mu} $ is the unit normal to the homogeneous hypersurface. It may be pointed out here that both 
$ G_{\mu \nu}  n^{\mu} n^{ \nu}$ and $R_{\mu \nu} n^{\mu} n^{ \nu} $ are
 expressed
in terms of the three geometry of the homogeneous hypersurfaces and the 
extrinsic curvature $K_{\mu \nu} = \bigtriangledown_{\nu} n_{\mu}$ respectively. The
 extrinsic curvature can be decomposed into its trace $K$ and trace-free part 
$\sigma_{\mu \nu}$ which represents the shear of the timelike geodesic congruence
orthogonal to the homogeneous hypersurface
\begin{equation}
K_{\mu \nu} = \frac{1}{3} K h_{\mu \nu} + \sigma_{\mu \nu}
\end{equation}
where $h_{\mu \nu} = g_{\mu \nu} + n_{\mu} n_{\nu} $,  projects orthogonal to 
$n_{\mu}$. We consider a scalar field theory to describe the 
energy momentum tensor which is given by
\begin{equation}
T_{\mu \nu} = \phi,_{\mu} \phi,_{\nu} - g_{\mu \nu} \left[ \frac{1}{2} g^{\alpha \beta} 
\phi,_{\alpha} \phi,_{\beta}  + V( \phi ) \right]
\end{equation}
with $V(\phi) = V_{o} + a \phi^{4}$, where $a$ and $V_{o}$ are constant. For
a homogeneous scalar field the dynamical equation (6) and (7) can  now be written
as
\begin{equation}
K^{2} =  3 \kappa^{2} \left( \frac{1}{2} \dot{\phi}^{2} + 
 V (\phi) \right) + \frac{ \tilde{\kappa}^{4}}{4} 
 \left( \frac{1}{2} \dot{\phi}^{2} + 
 V (\phi) \right)^{2} + \frac{3}{2} \sigma_{\mu \nu} \sigma^{\mu \nu}
- \frac{3}{2} \; ^{(3)}R - 3 {\it E}_{\mu \nu} n^{\mu} n^{\nu},
\end{equation}
\begin{equation}
\dot{K}  = \kappa^{2}  \left( - \dot{\phi}^{2} + 
 V (\phi) \right) -  \frac{\tilde{\kappa}^{4}}{4} 
 \left( \frac{1}{2} \dot{\phi}^{2} + 
 V (\phi) \right)  
 \left( \frac{5}{2} \dot{\phi}^{2} -  V (\phi) \right)  - \frac{1}{3} K^{2} - \sigma_{\alpha \beta} 
\sigma^{\alpha \beta}  + {\it E}_{\mu \nu } 
           n^{\mu} n^{\nu}
\end{equation}
and the wave equation for the scalar field is given by
\begin{equation}
\ddot{\phi} + 3 H \dot{\phi} = - \frac{dV}{d\phi}
\end{equation}
with $^{(3)}R$ as the scalar curvature of the homogeneous hypersurface.
In fact $^{(3)}R$ can be expressed in terms of the structure constants tensor 
$C_{\beta \mu}^{\alpha}$
of the Lie algebra of the spatial group ( of the space-time model) as
\begin{equation}
^{(3)}R = -  C_{\alpha \beta}^{\alpha} C^{\beta . \mu}_{. \mu} + \frac{1}{2}
C_{\beta \mu}^{\alpha} C^{\mu . \beta}_{. \alpha} - \frac{1}{4} C_{\alpha \beta \mu } C^{\alpha \beta \mu }
\end{equation}
( in this case raising and lowering of indices is done by $h_{\alpha \beta}$).
 Now, introducing the three form $\epsilon_{\alpha \beta \mu}$ 
( totally antisymmetric tensor) on the Lie algebra and using the 
antisymmetric property of the structure constants 
($C_{\alpha \beta  }^{\mu} = - C_{\beta \alpha}^{ \mu}$)
one can write [13]
\begin{equation}
C_{\alpha \beta  }^{\mu} =  M^{\mu \nu} \epsilon_{\nu \alpha \beta }
 + \delta^{\mu}_{[ \alpha} A_{\beta]}
\end{equation}
with $M^{\alpha \beta }$ as the symmetric tensor and $A_{\alpha}$ the dual vector. 
If we consider the Jacobi identity
\[
C^{\mu}_{\nu[\alpha} C^{\nu}_{\beta \mu ]} = 0
\]
then one gets
\begin{equation}
M^{\alpha \beta  } A_{\beta} = 0.
\end{equation}
Consequently one obtains the three  space curvature which is given by
\begin{equation}
^{(3)}R =  - \frac{3}{2} A_{\beta} A^{\beta} - h^{- 1} \left( M^{\alpha \beta }
 M_{\alpha \beta } - \frac{1}{2} M^{2}
\right)
\end{equation}
We note that $
^{(3)}R > 0$ only if $ M^{\alpha \beta }$ is either positive definite or negative. But from
equation (15) we have $A_{\beta} = 0$, which is the symmetry for Bianchi type IX spacetime.
Hence, except for Bianchi type IX we always get
\begin{equation}
^{(3)}R \leq 0
\end{equation}
Thus using the inequality it is evident from equation (10)
 that the constraint equation
leads to the inequality
$K^{2} > 0$ i.e., $K > 0$ (i.e., it will expand for ever) if the space-time is 
initially expanding and satisfying the constraint given by
\begin{equation}
E_{\mu \nu} n^{\mu} n^{\nu} \leq 0.
\end{equation}
So, from equation (10) we have
\begin{equation}
K^{2} >  3 \kappa^{2} \left[ \frac{1}{2} \dot{\phi}^{2} + 
 V (\phi) \right] + \frac{ \tilde{\kappa}^{4}}{4} 
 \left[ \frac{1}{2} \dot{\phi}^{2} + 
 V (\phi) \right]^{2} 
\end{equation}
for all time $t$.
Let us now define (using the ideas of  Wald's [13]  and the
 previous paper in GTR [15])
\[
K_{\phi} = K^{2} - 3 \kappa^{2} \left( \frac{1}{2} \dot{\phi}^{2} + 
 V (\phi) \right) + \frac{ \tilde{\kappa}^{4}}{4} 
 \left( \frac{1}{2} \dot{\phi}^{2} + 
 V (\phi) \right)^{2} 
\]
\begin{equation}
=  \frac{3}{2} \sigma_{\mu \nu} \sigma^{\mu \nu}
- \frac{3}{2} ^{(3)}R - 3 {\it E}_{\mu \nu} n^{\mu} n^{\nu}.
\end{equation}
Using the inequalities (17) and (18) it is found that
 $K_{\phi}$ is always a positive definite. 
The
constraint equation in Brane gets modified from that of  GTR [15], 
due to the presence of the term
quadratic in energy momentum.
The time differentiation of equation (20) can be written as
\begin{equation}
\dot{K}_{\phi}  = - \frac{2}{3} K K_{\phi} - 2 K \; \left( \sigma_{\mu \nu} 
 \sigma^{\mu \nu} \right)  + 2 K \left( {\it E}_{\mu \nu }  n^{\mu} n^{\nu} 
\right).
\end{equation}
Since we have  ${\it E}_{\mu \nu} n^{\mu} n^{\nu} \leq  0$, one obtains
\begin{equation}
\dot{K}_{\phi} \leq - \frac{2}{3} K K_{\phi}.
\end{equation}
Using the Hubble parameter ($H$) we rewrite the inequality as
\begin{equation}
\dot{K_{\phi}} \leq - 2 H K_{\phi}.
\end{equation}
On integrating the above inequality ( we note that $ K_{\phi} \geq 0$ ) one obtains
\begin{equation}
0 \leq K_{\phi} \leq  K_{\phi_{o}} e^{- 2 \int H dt} .
\end{equation}
where $K_{\phi_{o}}$ is an arbitrary constant.
During inflation ( exponential expansion ) $H$ can be assumed to be a constant ($= H_{o}$)
and we get
\begin{equation}
0 \leq K_{\phi} \leq  K_{\phi_{o}} e^{- 2 H_{o} t} .
\end{equation}
The expression $K_{\phi}$ falls of exponentially with time and 
goes to zero. However, if we consider  a case where ($H \sim \frac{H_{o}}{t}$)
then one obtains
\begin{equation}
0 \leq K_{\phi} \leq  K_{\phi_{o}} t^{- 2 H_{o} } .
\end{equation}
In this case $K_{\phi}$ decreases with time which follows a power law expansion
mode.
Thus the behavior of  $K_{\phi}$ is not affected by the nature of the potential.
In the brane-world scenario the quadratic term in energy density dominates. Thus
at very
high energy, we get
\begin{equation}
\dot{K} + \frac{1}{3} K^{2} \leq   \frac{ \tilde{\kappa}^{4}}{12} 
 \left( \frac{1}{2} \dot{\phi}^{2} + 
 V (\phi) \right)^{2} 
\end{equation}
which is different from that obtained in GTR [15]. During slow-roll inflation, the
scalar field decreases and the potential term behaves as a cosmological constant.
Consequently one can write equation (11) as
\begin{equation}
\dot{K}  + \frac{1}{3} K^{2} \leq  \kappa^{2} V_{o} \left[
1 +  \frac{V_{o}}{2  \lambda} \right].
\end{equation}
Now integrating the above inequality once we get
\begin{equation}
K \leq   \frac{ 3 \eta}{tanh \; (\eta t)} 
\end{equation}
where $\eta = \sqrt{ \frac{ \kappa^{2} V_{o}}{3}  \left( 1 + 
\frac{V_{o}}{2 \lambda}  \right)}$.
Thus $K$, the expansion rate, approches $3 \eta$ exponentially over
the time scale $\frac{1}{\eta}$. In the Brane world i.e., at very
 high energy scale ($\frac{V_{o}}{\lambda} \rightarrow \infty $) we have 
$\eta = \eta_{Brane} =  \frac{
\kappa V_{o}}{\sqrt{6  \lambda}} $ whereas in GTR i.e.,
 in the low energy scale i.e.,
 ($\frac{V_{o}}{\lambda} \rightarrow 0 $) it becomes $\eta = \eta_{GTR} = \kappa 
\sqrt{ \frac{V_{o}}{3}
}$. Thus $\frac{\eta_{Brane}}{\eta_{GTR}} = \sqrt{\frac{V_{o}}{2 \lambda}} $, which implies  that the time scale of approach of the inflationary regime 
in Brane world
is faster than that in GTR. 

Now using equation (29) as the upper limit of $K$ we have
\begin{equation}
\frac{1}{2} \sigma_{\mu \nu}  \sigma^{\mu \nu} \leq \frac{1}{3} \left( K^{2} - 9 \eta^{2} \right)
\leq \frac{3 \; \eta^{2}}{ Sinh^{2} (\eta \; t)}.
\end{equation}
Finally,  as $ K \rightarrow 3 \eta$, the shear $\sigma_{\mu \nu}$ approaches zero at late times. The time dependence of the spatial
metric can be approximately written as
\begin{equation}
h_{\mu \nu} (t) \sim e^{2 (t - t_{o}) \eta} \; h_{\mu \nu} (t_{o})
\end{equation}
where $t_{o}$ is the initial time.  The spatial
curvature $^{(3)}R$  scales away to zero. Thus the Bianchi universes (except Bianchi type
IX) after inflation appear to be matter free with nearly flat spatial sections for
time $t > \frac{1}{\eta}$ (i.e., isotropized) and the constant rate of isotropic
expansion is $K \rightarrow 3 \eta$.

To conclude, it is found that in the Brane world scenario, for all the Bianchi models ( except type IX ) it is possible to have either exponential or  power law inflation with a scalar field with arbitrary 
potential. It is also noted that in the brane world scenario
the universe  
isotropizes faster than  that in GTR. The potential considered here is $\phi^{4}$-type
with a constant additive part which  
influences the dynamics of the late
universe.

\vspace{0.5in}

{\large \it Acknowledgement :}
The work is supported by  the Minor Research Project grant by the University 
Grants Commission, New Delhi and North Bengal University.
I would like to  thank  
Prof. S. Mukherjee
for providing facilities of 
IUCAA Reference Centre at North Bengal University
where this work was carried out. 

\pagebreak

\end{document}